\documentclass[seceq]{ptptex}
\usepackage{wrapft}
\newtheorem{theorem}{Theorem}[section]

     \newenvironment{proof}[1][Proof.]{\begin{trivlist}
     \item[\hskip \labelsep {\bfseries #1}]}{\end{trivlist}}

     \newenvironment{remark}[1][Remark.]{\begin{trivlist}
     \item[\hskip \labelsep {\bfseries #1}]}{\end{trivlist}}
\newcommand{\qed}{\nobreak \ifvmode \relax \else
      \ifdim\lastskip<1.5em \hskip-\lastskip
      \hskip1.5em plus0em minus0.5em \fi \nobreak
      \vrule height0.75em width0.5em depth0.25em\fi}
\notypesetlogo  %comment in if to eliminate PTPTeX logo
\title{%        %You can use \\ for explicit line-break
The Dirac propagator in the extreme Kerr metric
}

\author{%       %Use \scshape  for the family name
Davide \textsc{Batic}$^{1,}$\footnote{E-mail: batic@itp.phys.ethz.ch} 
and Harald \textsc{Schmid}$^{2,}$\footnote{E-mail: Harald.Schmid@UBH.de}
}

\inst{%         %Affiliation, neglected when [addenda] or [errata]
$^1$Institute for Theoretical Physics, ETH Z\"{u}rich, CH-8093 Z\"{u}rich, Switzerland\\
$^2$UBH Software \& Engineering GmbH, D-92224 Amberg, Germany}

\abst{%         %this abstract is neglected when [addenda] or [errata]
Starting with the Dirac equation in the extreme Kerr metric we derive an integral representation for the propagator of solutions of the Cauchy problem with initial data in the class of smooth compactly supported functions.
}

\begin{document}

\maketitle

\section{Introduction}\label{sec:1}
One may think that the formation of an extreme Kerr black hole (EKBH) is of only academic interest but this is not so. In 2003 it has been proved that the so-called relativistic Dyson rings, i.e. uniformly rotating, homogeneous and axisymmetric relativistic fluid bodies with a toroidal shape admit a continuous transition to an EKBH if a fixed ratio $r_1/r_2>0.5613$ of inner to outer coordinate radius is prescribed and the gravitational mass gradually increases for fixed mass-density.\cite{ANS} Concerning the existence of such rings based on numerical computations we refer to Ref.~\citen{WON}-\citen{ERI}. Let us recall that relativistic Dyson rings could emerge from astrophysical scenarios like stellar core-collapses with high angular momentum\cite{TH} or they might simply be present in central regions of galaxies. Moreover, in Ref.~\citen{MEI} it has been proved that the only possible candidate for a black hole limit for stationary and axisymmetric, uniformly rotating perfect fluid bodies with a cold equation of state as well as for isentropic stellar models with a non-zero temperature is the EKBH. Hence, we cannot a priori exclude that EKBH's play no role in Astrophysics and in view of the above consideration it is reasonable to study the Dirac equation in an extreme Kerr manifold.\\
It is surprising that there are no analytical studies concerning the propagation of Dirac fields outside an extreme Kerr black hole. With the present work we hope to fill this gap. Here, we prove for the first time the completeness of the Chandrasekhar ansatz for the Dirac equation in an EKBH from which an integral representation of the Dirac propagator can be obtained quite immediately.\\
The rest of the paper is organized as follows. In Section~\ref{sec:2} we shortly derive the Dirac Equation in the EKBH. After the introduction of the so-called Chandrasekhar ansatz we compute the scalar product with respect to which the Dirac Hamiltonian is formally self-adjoint. Section~\ref{sec:3} is devoted to prove the completeness of the Chandrasekhar ansatz (see Theorem~\ref{completo}) which in turn allows us to derive the integral representation for the Dirac propagator as given by \eqref{propagator}.

\section{The Dirac equation in the extreme Kerr metric}\label{sec:2}
In Boyer-Lindquist coordinates $(t,r,\vartheta,\varphi)$ with $r>0$, $0\leq\vartheta\leq\pi$, $0\leq\varphi<2\pi$ the extreme Kerr metric can be easily derived from the Kerr metric\cite{WAL} by setting the Kerr parameter $a=M$. Its form is given by
\begin{multline}\label{KN}
ds^{2}=\left(1-\frac{2Mr}{\Sigma}\right)dt^{2}+\frac{4M^{2}r\sin^{2}\vartheta}{\Sigma}dtd\varphi-\frac{\Sigma}{\Delta}dr^2-\Sigma d\vartheta^{2}\\
-(r^2+M^2)^{2}\sin^{2}{\vartheta}\frac{\widetilde{\Sigma}}{\Sigma}d\varphi^{2}
\end{multline}
with
\[
\Sigma:=\Sigma(r,\theta)=r^2+M^2\cos^{2}\theta, \qquad \Delta:=\Delta(r)=(r-M)^2
\]
and
\[
\widetilde{\Sigma}:=\widetilde{\Sigma}(r,\vartheta)=1-M^{2}\gamma^{2}(r)\sin^{2}{\vartheta},\quad \gamma(r):=\frac{r-M}{r^2+M^2}
\]
where $M$ is the mass of a spinning black hole with angular momentum $J=M^2$. Notice that the area of an EKBH is simply $A=8\pi J$. Since the equation $\Delta=0$ has a double root at $r_{0}:=M$ the Cauchy horizon and the event horizon coincide. Finally, notice that $\widetilde{\Sigma}>0$ for all $r>M$ and $\vartheta\in[0,\pi]$.\\
According to Penrose and Rindler \cite{PEN1} the Dirac equation coupled to a general gravitational field $\textbf{V}$ is given in terms of two-component spinors $(\phi^{A},\chi^{A^{'}})$ by 
\[
(\nabla^{A}_{A^{'}}-i eV^{A}_{A^{'}})\phi_{A}=\frac{m_{e}}{\sqrt{2}}\chi_{A^{'}},\qquad   
(\nabla_{A}^{A^{'}}-i eV_{A}^{A^{'}})\chi_{A^{'}}=\frac{m_{e}}{\sqrt{2}}\phi_{A}
\]
where we used Planck units $\hbar=c=G=1$. Furthermore, $\nabla_{AA^{'}}$ is the symbol for covariant differentiation, $e$ is the charge or coupling constant of the Dirac particle to the vector field $\textbf{V}$ and $m_{e}$ is the particle mass.
The Dirac equation in the Kerr geometry was computed and separated by Chandrasekhar \cite{CH1} with the help of the Kinnersley tetrad \cite{KIN}. Since the derivation and separation of the Dirac equation in the extreme Kerr geometry follows with minor changes from $\S 2$ in Batic and Schmid\cite{DH} we will limit us to give here only the main results. In view of the separation of the Dirac equation we choose to work with the Carter tetrad \cite{CA1}. A Dirac spinor $\psi=\psi(t,r,\vartheta,\varphi)\in\mathbb{C}^{4}$ satisfies in the exterior region $r\in(M,\infty)$ of an extreme Kerr black hole the following equation
\begin{equation} \label{Dirac}
\mathcal{W}\psi=\left(\mathcal{W}_{(t,r,\varphi)}+\mathcal{W}_{(t,\vartheta,\varphi)}\right)\psi=0
\end{equation}
where
\begin{equation} \label{2}
\mathcal{W}_{(t,r,\varphi)}=\left( \begin{array}{cccc}
                            im_{e}r&0&\sqrt{\Delta}\mathcal{D}_{+}&0\\
                            0&-im_{e}r&0&\sqrt{\Delta}\mathcal{D}_{-}\\
                            \sqrt{\Delta}\mathcal{D}_{-}&0&-im_{e}r&0\\
                             0&\sqrt{\Delta}\mathcal{D}_{+}&0&im_{e}r
                            \end{array} \right),
\end{equation}
\begin{equation} \label{3}
\mathcal{W}_{(t,\vartheta,\varphi)}=\left( \begin{array}{cccc}
                            -Mm_{e}\cos{\vartheta}&0&0&\mathcal{L}_{+}\\
                            0&Mm_{e}\cos{\vartheta}&-\mathcal{L}_{-}&0\\
                            0&\mathcal{L}_{+}&-Mm_{e}\cos{\vartheta}&0\\
                            -\mathcal{L}_{-}&0&0&Mm_{e}\cos{\vartheta}
                            \end{array} \right)
\end{equation}
with $\mathcal{D}_{\pm}$ and $\mathcal{L}_{\pm}$ defined by
\begin{eqnarray}
&&\mathcal{D}_{\pm}=\frac{\partial}{\partial r}\mp\frac{1}{\Delta}\left[(r^2+M^2)\frac{\partial}{\partial t}+M\frac{\partial}{\partial\varphi}\right],\label{DPM}\\
&&\mathcal{L}_{\pm}=\frac{\partial}{\partial \vartheta}+\frac{1}{2}\cot{\vartheta}\mp i\left(M\sin{\vartheta}\frac{\partial}{\partial t}+\csc{\vartheta}\frac{\partial}{\partial\varphi}\right)\label{LPM}.
\end{eqnarray}
If we make the Chandrasekhar ansatz\cite{CH1,DH}
\[
\psi(t,r,\vartheta,\varphi)=e^{i\omega t}e^{i\kappa\varphi}\left( \begin{array}{c}
               R_{-}(r)S_{-}(\vartheta)\\
               R_{+}(r)S_{+}(\vartheta)\\
               R_{+}(r)S_{-}(\vartheta)\\
               R_{-}(r)S_{+}(\vartheta)
\end{array} \right),\quad\kappa=k+\frac{1}{2}
\]
with $\omega$ the energy of the particle and $k\in\mathbb{Z}$ its azimuthal quantum number, then the Dirac equation decouples into the following systems of linear first order differential equations for the radial $R_{\pm}$ and angular components $S_{\pm}$ of the spinor $\psi$ 
\begin{eqnarray} 
&&\left( \begin{array}{cc}
     \sqrt{\Delta}~\widehat{\mathcal{D}}_{-}&-i m_{e}r-\lambda\\
     i m_{e}r-\lambda&\sqrt{\Delta}~\widehat{\mathcal{D}}_{+}
           \end{array} \right)\left( \begin{array}{cc}
                                     R_{-} \\
                                     R_{+}
                                     \end{array}\right)=0, \label{radial}\\
&&\left( \begin{array}{cc}
     -\widehat{\mathcal{L}}_{-} & \lambda+Mm_{e}\cos\theta\\
                \lambda-Mm_{e}\cos\theta & \widehat{\mathcal{L}}_{+}
           \end{array} \right)\left( \begin{array}{cc}
                                     S_{-} \\
                                     S_{+}
                                     \end{array}\right)=0 \label{angular}
\end{eqnarray} 
where
\begin{eqnarray}
&&\widehat{\mathcal{D}}_{\pm}=\frac{d}{dr}\mp i\frac{K(r)}{\Delta},\quad \hspace{1.5cm}K(r)=\omega(r^2+M^2)+\kappa M, \label{ramb1}\\
&&\widehat{\mathcal{L}}_{\pm}=\frac{d}{d\vartheta}+\frac{1}{2}\cot{\vartheta}\pm Q(\vartheta),\quad Q(\vartheta)=M\omega\sin{\vartheta}+\kappa\csc{\vartheta} \label{ramb2}
\end{eqnarray}
and $\lambda$ is a separation constant depending on $k$ and $\omega$. Let $u\in\mathbb{R}$ be the tortoise coordinate defined by $du/dr=(r^{2}+M^{2})/\Delta$. By rearranging \eqref{Dirac} we can write the Dirac equation in Hamiltonian form
\begin{equation}\label{Dirac_Ham}
i\frac{\partial\psi}{\partial t}=H\psi,\quad H=H_{0}+V(u,\vartheta)
\end{equation}
with
\begin{equation} \label{hamilt4}
H_{0}=A(u,\vartheta)\left[\left( \begin{array}{cccc}
                            -\mathcal{E}_{-}&0&0&0\\
                            0&\mathcal{E}_{+}&0&0\\
                            0&0&\mathcal{E}_{+}&0\\
                            0&0&0&-\mathcal{E}_{-}
         \end{array} \right)+\left( \begin{array}{cccc}
                            0&-\mathcal{M}_{+}&0&0\\
                            -\mathcal{M}_{-}&0&0&0\\
                            0&0&0&\mathcal{M}_{+}\\
                            0&0&\mathcal{M}_{-}&0
         \end{array} \right)\right],
\end{equation}
and
\begin{eqnarray}
&&A(u,\vartheta)=\frac{1}{\widetilde{\Sigma}}\left[1\hspace{-1mm}\rm{I}_{4}-\it{M\gamma(u)}\sin{\vartheta}\left(\begin{array}{cc}
       \sigma_{2}&0\\
       0&-\sigma_{2}
           \end{array} \right)\right], \label{matrixA}\\
&&V(u,\vartheta)=m_{e}A(u,\vartheta)\gamma(u)
\left(\begin{array}{cc}
       0&\frac{1\hspace{-1mm}\rm{I}_{2}}{\widetilde{\rho}} \\
       \frac{1\hspace{-1mm}\rm{I}_{2}}{\overline{\widetilde{\rho}}}&0
           \end{array}\right) \label{potential}
\end{eqnarray}
where $\sigma_{2}$ is a Pauli matrix, $\widetilde{\rho}=-(r-iM\cos{\vartheta})^{-1}$ and
\[
\mathcal{E}_{\pm}=i\frac{\partial}{\partial u}\mp\frac{iM}{r^2+M^2}\frac{\partial}{\partial\varphi},\quad\quad\mathcal{M}_{\pm}=i\gamma(u)\left(\frac{\partial}{\partial\vartheta}+\frac{1}{2}\cot{\vartheta}\mp i\csc{\vartheta}\frac{\partial}{\partial\varphi}\right)
\]
satisfying $\overline{\mathcal{E}}_{\pm}=-\mathcal{E}_{\pm}$ and $\overline{\mathcal{M}}_{\pm}=-\mathcal{M}_{\mp}$. Notice that the matrix contained in the square brackets in \eqref{matrixA} is hermitian. Similarly as in $\S 3$ in Batic and Schmid\cite{DH} we can construct a positive scalar product 
\begin{equation}\label{scalar_product}
\langle\psi|\phi\rangle=\int_{-\infty}^{+\infty}\,du\int_{-1}^{1}\,d(\cos{\vartheta})\int_{0}^{2\pi}\,d\varphi\sqrt{\widetilde{\Sigma}}~\overline{\psi}\phi
\end{equation}
with respect to which the Hamiltonian $H$ acting on the spinor $\psi$ on the hypersurface $t=$const. is formally self-adjoint. In the following we consider the Hilbert space $\mathcal{H}=L_{2}(\Omega)^{4}:=L_{2}\left(\Omega,\sqrt{\widetilde{\Sigma}}\hspace{1mm}du\hspace{1mm}d(\cos{\vartheta})\hspace{1mm}d\varphi\right)^{4}$ consisting of wave functions $\psi:\Omega:=\mathbb{R}\times S^{2}\to\mathbb{C}^{4}$ together with the positive scalar product \eqref{scalar_product}. Similarly as in Ref.~\citen{DH} it can be shown that the Hamiltonian $H$ defined on $\mathcal{C}^{\infty}_{0}(\Omega)^4$ is essentially self-adjoint and has a unique self-adjoint extension. We recall now some few basic facts from Ref.~\citen{DH}. The system \eqref{angular} can be brought in the so-called Dirac form
\[
(\mathcal{U}S)(\vartheta):=\left( \begin{array}{cc}
       0&1\\
      -1&0
                     \end{array} \right)S^{'}+\left( \begin{array}{cc}
       -Mm_{e}\cos{\vartheta}&-\frac{\kappa}{\sin{\vartheta}}-M\omega\sin{\vartheta}\\
      -\frac{\kappa}{\sin{\vartheta}}-M\omega\sin{\vartheta}&Mm_{e}\cos{\vartheta}
                     \end{array} \right)S=\lambda S
\]
with $S(\vartheta)=\sqrt{\sin{\vartheta}}(S_{-}(\vartheta),S_{+}(\vartheta))^{T}$ and $\vartheta\in(0,\pi)$. In $L_{2}(0,\pi)^{2}$ the angular operator $\mathcal{U}$ with domain $D(\mathcal{U})=\mathcal{C}_{0}^{\infty}(0,\pi)^{2}$ is essentially self-adjoint, its spectrum is discrete and consists of simple eigenvalues, i.e. $\lambda_{j}<\lambda_{j+1}$ for every $j\in\mathbb{Z}\backslash\{0\}$. Moreover, the eigenvalues depend smoothly on $\omega$. Furthermore, the functions $e^{i\left(k+\frac{1}{2}\right)\varphi}$ are eigenfunctions of the z-component of the total angular momentum operator $\widehat{J}_{z}$ with eigenvalues $-\left(k+\frac{1}{2}\right)$ with $k\in\mathbb{Z}$. Since energy, generalized squared total angular momentum and the z-component of the total angular momentum form a set of commuting observables $\{H,\widehat{J}^{2},\widehat{J}_{z}\}$ we will label the generalized states $\psi\in\mathcal{H}$ by $\psi^{kj}_{\omega}$. Finally, for every $k\in\mathbb{Z}$ and $j\in\mathbb{Z}\backslash\{0\}$ the set $\{Y^{kj}_{\omega}(\vartheta,\varphi)\}$ with 
\begin{equation} \label{ypsilon}
Y^{kj}_{\omega}(\vartheta,\varphi)=\left( \begin{array}{c}
      Y_{\omega,-}^{kj}(\vartheta,\varphi) \\
      Y_{\omega,+}^{kj}(\vartheta,\varphi)
           \end{array} \right)=\frac{1}{\sqrt{2\pi}}\left( \begin{array}{c}
      S_{\omega,-}^{kj}(\vartheta) \\
      S_{\omega,+}^{kj}(\vartheta)
           \end{array} \right)e^{i\left(k+\frac{1}{2}\right)\varphi},
\end{equation}
is a complete orthonormal basis for $L_{2}(S^{2})^{2}$. Let $\sigma(H)\subseteq\mathbb{R}$ denote the spectrum of the self-adjoint Hamiltonian operator $H$. We state now the theorem on the completeness of the Chandrasekhar ansatz.
\begin{theorem} \label{completo}
For every $\psi\in\mathcal{C}_{0}^{\infty}(\Omega)^{4}$
\begin{equation} \label{completeness}
\psi(0,x)=\int_{\sigma(H)}\,\sum_{j\in\mathbb{Z}\backslash\{0\}}\sum_{k\in\mathbb{Z}}\langle\psi^{kj}_{\omega}|\psi_{\omega}\rangle\psi^{kj}_{\omega}(x)d\mu_{\omega},\hspace{-2mm}\quad \psi^{kj}_{\omega}(x)=\left( \begin{array}{c}
               R_{\omega,-}^{kj}(u)Y_{\omega,-}^{kj}(\vartheta,\varphi)\\
               R_{\omega,+}^{kj}(u)Y_{\omega,+}^{kj}(\vartheta,\varphi)\\
               R_{\omega,+}^{kj}(u)Y_{\omega,-}^{kj}(\vartheta,\varphi)\\
               R_{\omega,-}^{kj}(u)Y_{\omega,+}^{kj}(\vartheta,\varphi)
\end{array} \right)
\end{equation}
where the scalar product $\langle\cdot|\cdot\rangle$ is given by \eqref{scalar_product}, $\mu_{\omega}$ is a Borel measure on $\sigma(H)\subseteq\mathbb{R}$ and $x=(u,\vartheta,\varphi)$.  
\end{theorem}
\begin{proof}
Similarly as in Thm.~5.2 in Ref.~\citen{DH} we construct isometric operators
\[
\widehat{W}_{k,j}:\mathcal{C}_{0}^{\infty}(\mathbb{R})^{2}\longrightarrow \mathcal{C}_{0}^{\infty}(\Omega)^{4},
\]
such that 
\[
R^{kj}_{\omega}(u)=\left( \begin{array}{c}
               R^{kj}_{\omega,-}(u)\\
               R^{kj}_{\omega,+}(u)
\end{array} \right)\longmapsto(\widetilde{\Sigma})^{-\frac{1}{4}}\left( \begin{array}{c}
               R_{\omega,-}^{kj}(u)Y_{\omega,-}^{kj}(\vartheta,\varphi)\\
               R_{\omega,+}^{kj}(u)Y_{\omega,+}^{kj}(\vartheta,\varphi)\\
               R_{\omega,+}^{kj}(u)Y_{\omega,-}^{kj}(\vartheta,\varphi)\\
               R_{\omega,-}^{kj}(u)Y_{\omega,+}^{kj}(\vartheta,\varphi)
\end{array} \right).
\]
By means of the isometric operators $\widehat{W}_{k,j}$ we can now introduce for every $\omega\in\sigma(H)$ an auxiliary separable Hilbert space $\mathfrak{h}(\omega)$ as follows
\[
\mathfrak{h}(\omega)=\bigoplus_{j\in\mathbb{Z}\backslash{0}}\bigoplus_{k\in\mathbb{Z}}\mathfrak{h}_{k,j},\quad\mathfrak{h}_{k,j}=\widehat{W}_{k,j}(\mathcal{C}_{0}^{\infty}(\mathbb{R})^{2}).
\]
According to the expansion theorem (e.g. Th.3.7 in Ref.~\citen{WE2}) every element $\psi_{\omega}$ in $\mathfrak{h}(\omega)$ can be written as
\begin{equation} \label{opsala}
\psi_{\omega}=\sum_{j\in\mathbb{Z}\backslash{0}}\sum_{k\in\mathbb{Z}}\langle\psi^{kj}_{\omega}|\psi_{\omega}\rangle\psi^{kj}_{\omega}.
\end{equation}
Finally, the direct integral of Hilbert spaces
\begin{equation} \label{primdir}
\mathfrak{H}=\int_{\sigma(H)}\,\bigoplus_{j\in\mathbb{Z}\backslash{0}}\bigoplus_{k\in\mathbb{Z}}\mathfrak{h}_{k,j}d\mu_{\omega}
\end{equation}
with $\mu_{\omega}$ a Borel measure on $\sigma(H)\subseteq\mathbb{R}$ is defined (see Ch.1 $\S$5.1 in Ref.~\citen{YAF}) as the Hilbert space of vector valued functions $\psi_{\omega}$ taking values in the auxiliary Hilbert spaces $\mathfrak{h}(\omega)$. By definition $\mathcal{H}$ can be written as in \eqref{primdir} if there exists a unitary mapping $\mathcal{F}$ of $\mathcal{H}$ onto $\mathfrak{H}$. Now, since the Hamiltonian $H$ is self-adjoint, the spectral representation theorem (e.g. Theorem 7.18 in Ref.~\citen{WE2}) implies the existence of such an isomorphism $\mathcal{F}$ and this completes the proof.\qed
\end{proof}
At this point notice that in general for self-adjoint operators\cite{KALF}
\[
\sigma(H)=\sigma_{p}(H)\cup\sigma_{c}(H)
\]
where $\sigma_{p}(H)$ and $\sigma_{c}(H)$ denote the point spectrum and the continuous spectrum, respectively. Furthermore, we distinguish between points of $\sigma_{p}(H)$ which are isolated or non-isolated as points of $\sigma(H)$. The former constitutes the discrete spectrum $\sigma_{d}(H)$ which is defined as the subset of $\sigma(H)$ for which the resolvent is closed. The latter is called the point continuous spectrum $\sigma_{pc}(H)$. In the next section we investigate the spectrum of the Hamiltonian $H$ in order to compute the measure $\mu_{\omega}$ entering in \eqref{completeness}.
\section{The Dirac Propagator}\label{sec:3}
According to Theorem~\ref{completo} every representative element $\psi_{\omega}=(\mathcal{F}\psi)(\omega)$ of the element $\psi\in\mathcal{H}$ in the decomposition (\ref{primdir}) can be written in the form given by (\ref{opsala}). Hence, in the study of $\sigma(H)$ we are allowed to focus our analysis on the radial system (\ref{radial}). As a consequence we just need to investigate the spectrum of the differential operator $\mathcal{R}$ associated to the formal differential system (\ref{radial}) after it is brought into the form of a Dirac system of ordinary differential equations. This can be achieved in two steps. First, we set $R_{-}(r)=F(r)+iG(r)$ and $R_{+}(r)=F(r)-iG(r)$ in \eqref{radial} and we obtain the following system
\begin{eqnarray*}
\frac{dF}{dr}&=&\frac{\lambda}{r-M}F+\left[\frac{K(r)}{(r-M)^2}+\frac{m_{e}r}{r-M}\right]G,\\
\frac{dG}{dr}&=&-\frac{\lambda}{r-M}G+\left[-\frac{K(r)}{(r-M)^2}+\frac{m_{e}r}{r-M}\right]F.
\end{eqnarray*}
By introducing the tortoise coordinate $u\in\mathbb{R}$ defined in Section~\ref{sec:2} the above equations give the following first order system, namely
 \begin{equation}\label{Dirac_system_0}
\Xi^{'}=
\left(\begin{array}{cc}
     \lambda\Delta\Omega&\omega+\kappa\Omega+m_{e}r(u)\Delta\Omega\\
     -\omega-\kappa\Omega+m_{e}r(u)\Delta\Omega&-\lambda\Delta\Omega
\end{array}\right)\Xi, \quad u\in\mathbb{R}
\end{equation}
with $^{'}:=d/du$, $\Xi:=(F,G)^{T}$ and
\[
\Omega:=\Omega(u)=\frac{M}{r(u)^2+M^2},\quad\Delta\Omega:=\Delta\Omega(u):=\frac{r(u)}{r^2(u)+M^2}-\Omega(u).
\]
For ease in notation we shall write $r$ instead of $r(u)$ when no risk of confusion arises. Finally, we rewrite \eqref{Dirac_system_0} as follows
\begin{equation}\label{Dirac_system}
\left(\begin{array}{cc}
     0&-1\\
    1&0
\end{array}\right)\Xi^{'}+
\left(\begin{array}{cc}
     -\kappa\Omega-m_{e}r\Delta\Omega&-\lambda\Delta\Omega\\
    -\lambda\Delta\Omega&-\kappa\Omega+m_{e}r\Delta\Omega
\end{array}\right)\Xi=\omega\Xi, \quad u\in\mathbb{R}.
\end{equation}
Notice that the value of $\Omega$ at the event horizon is simply the black hole angular velocity $\Omega_{H}=1/(2M)$. Moreover, $\Delta\Omega\to 0$ for $u\to-\infty$. Let us now consider the formal differential operator
\begin{equation}\label{tau}
\tau:=\left(\begin{array}{cc}
     0&-1\\
    1&0
\end{array}\right)\frac{d}{du}-
\left(\begin{array}{cc}
     \kappa\Omega+m_{e}r\Delta\Omega&\lambda\Delta\Omega\\
     \lambda\Delta\Omega&\kappa\Omega-m_{e}r\Delta\Omega
\end{array}\right)
\end{equation}
in the Hilbert space $L_{2}(\mathbb{R},du)^2$. Thm.~6.8 in Ref.~\citen{WE1} implies that $\tau$ is in the limit point case (l.p.c.) at $\pm\infty$. Hence, the deficiency indices $(\gamma_{+},\gamma_{-})$ of $\tau$ are $(0,0)$, its deficiency numbers are $(\gamma_{-\infty},\gamma_{\infty})=(1,1)$ and $\tau$ admits only one self-adjoint extension $T$ with $D(T)=\mathcal{C}^{\infty}_{0}(\mathbb{R})^2$ (see Ch.~4 p.~53 and Thm.~5.7-8 in Ref.~\citen{WE1}).
\begin{remark}
\textit{The existence of a unique self-adjoint extension ensures that there is no need to choose some boundary conditions for the Dirac particle on the manifold under consideration.}
\end{remark}
The result stated in the next theorem is a particular case of a more general result obtained in Ref.~\citen{BEL}.
\begin{theorem}\label{essential}
Let $T$ with $D(T)=\mathcal{C}^{\infty}_{0}(\mathbb{R})^2$ be the self-adjoint extension of the formal differential operator $\tau$ defined in \eqref{tau}. Then, $\sigma_{e}(T)=\mathbb{R}=\sigma(T)$ where $\sigma_{e}(\cdot)$ denotes the essential spectrum.
\end{theorem}
\begin{proof}
In order to determine the essential spectrum of $T$ we apply the so-called decomposition method due to Neumark (see Ref.~\citen{NEU}). To this purpose let $T_{-}$ and $T_{+}$ be self-adjoint extensions of the operator $T$ restricted to the intervals $(-\infty,0]$ and $[0,\infty)$, respectively. Moreover, let the $2\times 2$ symmetric matrix $P(u)$ be defined as follows
\[
P(u):=\left(\begin{array}{cc}
     -\kappa\Omega-m_{e}r\Delta\Omega&-\lambda\Delta\Omega\\
    -\lambda\Delta\Omega&-\kappa\Omega+m_{e}r\Delta\Omega
\end{array}\right).
\]  
A straightforward computation shows that
\[
P_{0}:=\lim_{u\to +\infty}P(u)=\left(\begin{array}{cc}
    -m_{e}&0\\
    0&m_{e}
\end{array}\right),\quad P_{1}:=\lim_{u\to -\infty}P(u)=\left(\begin{array}{cc}
     -\kappa\Omega_{H}&0\\
    0&-\kappa\Omega_{H}
\end{array}\right).
\]
Let $\mu_{\pm}^{(i)}$ with $i=0,1$ denote the eigenvalues of $P_{0}$ and $P_{1}$, respectively.
Since $\mu_{-}^{(0)}=-m_{e}$, $\mu_{+}^{(0)}=m_{e}$ and $\mu_{-}^{(1)}=\mu_{+}^{(1)}=-\kappa\Omega_{H}=-\kappa/(2M)$ Thm.~16.5 in Ref.~\citen{WE1} implies that
\[
\sigma_{e}(T_{+})\cap(-m_{e},m_{e})=\emptyset,\quad \sigma_{e}(T_{-})\cap\emptyset=\emptyset. 
\]
Notice that the applicability of Thm.~16.5 continues to hold also for the case $m_{e}=0$. Let $d\in(0,\infty)$. A simple computation shows that for $u\to +\infty$
\[
P(u)-P_{0}=\left(\begin{array}{cc}
     m_{e}M&-\lambda\\
    -\lambda&-m_{e}M
\end{array}\right)\frac{1}{u}+\mathcal{O}\left(\frac{1}{u^2}\right).
\]
Hence, it follows that
\[
\lim_{x\to +\infty}\frac{1}{x}\int_{d}^{x}du\,|P(u)-P_{0}|=0
\]
and Thm. 16.6 in Ref.~\citen{WE1} implies that
\[
\sigma_{e}(T_{+})\supset(-\infty,m_{e}]\cup[m_{e},\infty).
\]
Let $\delta\in(-\infty,0)$. In the limit $r\to M$ we have
\begin{eqnarray*}
P(r)-P_{1}&=&\frac{1}{M}\left(\begin{array}{cc}
     \Omega_{H}(\kappa-m_{e}M)&-\lambda/(2M)\\
    -\lambda/(2M)&\Omega_{H}(\kappa+m_{e}M)
\end{array}\right)(r-M)+\mathcal{O}\left((r-M)^2\right),\\
\frac{du}{dr}&=&1+\frac{2M}{r-M}+\frac{2M^2}{(r-M)^2},\quad \frac{1}{u}=-\frac{1}{2M^2}(r-M)+\mathcal{O}\left((r-M)^2\right).
\end{eqnarray*}
Now, it can be easily checked that
\[
\lim_{y\to -\infty}\frac{1}{y}\int^{\delta}_{y}du\,|P(u)-P_{1}|=\lim_{r\to M}\frac{1}{y(r)}\int^{\widetilde{\delta}}_{r}d\widetilde{r}\,\frac{du}{d\widetilde{r}}|P(\widetilde{r})-P_{1}|=0
\]
and Thm.~16.6 implies that $\sigma_{e}(T_{-})\supset\mathbb{R}$. Finally, the remark to Thm. 11.5 in Ref.~\citen{WE1} yields that
\[
\sigma_{e}(T)=\sigma_{e}(T_{-})\cup\sigma_{e}(T_{+})=\mathbb{R}.\qed
\]
\end{proof}
From Theorem~\ref{essential} it results that $\sigma_{e}(H)=\mathbb{R}=\sigma(H)$. Since for the discrete spectrum $\sigma_{d}(H)=\sigma(H)\backslash\sigma_{e}(H)$ we find that $\sigma_{d}(H)=\emptyset$. Hence, the absolutely continuous spectrum of $H$ is simply $\sigma_{ac}(H)=\mathbb{R}$ and the purely point spectrum is empty, i.e. $\sigma_{pp}(H)=\emptyset$. On the other hand Schmid proved in Ref.~\citen{SC} that the Dirac Hamiltonian in the extreme Kerr metric admits an eigenvalue $\Omega_k=-\kappa\Omega_{H}$ for $|\omega|<m_{e}$ and for each $\kappa=k+1/2$ with $k\in\mathbb{Z}$. Hence, the point continuous spectrum $\sigma_{pc}(H)$ is not empty since $\sigma_{pc}(H)=\{\Omega_{k}\}$ for $k$ fixed. Thus we can conclude that such an eigenvalue is embedded in the continuous spectrum of $H$. Moreover, the Lebesgue's decomposition theorem leads to a decomposition of the spectral measure $\mu$ into the sum of a part absolutely continuous with respect to Lebesgue measure and a singular part, i.e.
\[
\mu=\mu_{ac}+\mu_{s}
\]
and the Radon-Nikodym theorem implies that the absolutely continuous part $\mu_{ac,\omega}$ of the spectral measure may be described by 
\[
\mu_{ac,\omega}=\int_{\mathbb{R}}d\omega\,f(\omega)
\]
with $f(\omega)$ a density function defined as $f(\omega)=d\rho(\omega)/d\omega$ for almost all $\omega\in\mathbb{R}$ and $\rho(\omega)$ the spectral function. Since the spectral measure of the operator $H$ coincides with the Lebesgue measure on the spectrum of $H$ (see Thm.~3.1 p.447 and Ch.VI$\S$5 in Ref.~\citen{BER}) the singular component $\mu_{s,\omega}$ is supported on the set of eigenvalues of the operator $H$. These may be characterized  as the points of discontinuity of the spectral function $\rho(\omega)$, i.e. points where $\rho(\omega)$ has an isolated jump. Without loss of generality we can choose $\mu_{s,\omega}=\widetilde{H}(\omega-\Omega_{k})$ where $\widetilde{H}$ is the Heaviside function. Taking into account that in the distributional sense the derivative of the Heaviside function gives rise to a Dirac-delta, we find that \eqref{completeness} becomes
\begin{equation}\label{completeness2}
\psi(x)=\int_{\mathbb{R}}d\omega\,\sum_{j\in\mathbb{Z}\backslash\{0\}}\sum_{k\in\mathbb{Z}}\langle\psi^{kj}_{\omega}|\psi_{\omega}\rangle\psi^{kj}_{\omega}(x)+\sum_{j\in\mathbb{Z}\backslash\{0\}}\sum_{k\in\mathbb{Z}}\langle\psi^{kj}_{\Omega_{k}}|\psi_{\Omega_k}\rangle\psi^{kj}_{\Omega_k}(x)
\end{equation}
with
\[
\psi^{kj}_{\omega}(x)=\left( \begin{array}{c}
               R_{\omega,-}^{kj}(u)Y_{\omega,-}^{kj}(\vartheta,\varphi)\\
               R_{\omega,+}^{kj}(u)Y_{\omega,+}^{kj}(\vartheta,\varphi)\\
               R_{\omega,+}^{kj}(u)Y_{\omega,-}^{kj}(\vartheta,\varphi)\\
               R_{\omega,-}^{kj}(u)Y_{\omega,+}^{kj}(\vartheta,\varphi)
\end{array} \right),\quad
\psi^{kj}_{\Omega_{k}}(x)=\left( \begin{array}{c}
               R_{\Omega_{k},-}^{kj}(u)Y_{\Omega_{k},-}^{kj}(\vartheta,\varphi)\\
               R_{\Omega_k,+}^{kj}(u)Y_{\Omega_k,+}^{kj}(\vartheta,\varphi)\\
               R_{\omega_k,+}^{kj}(u)Y_{\omega_k,-}^{kj}(\vartheta,\varphi)\\
               R_{\omega_k,-}^{kj}(u)Y_{\omega_k,+}^{kj}(\vartheta,\varphi)
\end{array} \right)
\]
where $R_{\Omega_k,\pm}^{kj}$ and $Y_{\Omega_k,\pm}^{kj}$  are the radial and angular eigenfunctions satisfying \eqref{radial} and \eqref{angular} with $\omega=\Omega_{k}$, respectively. In Thm.~3.6 (see Ref.~\citen{SC}) it has been shown that the radial eigenfunctions can be written in terms of Whittaker functions, which in turn can be expressed by means of generalized Laguerre polynomials whenever the condition $1+\alpha+\kappa+n=0$ is satisfied with $n\in\mathbb{N}$ and
\[
\alpha:=\frac{M(m^{2}_{e}-2\omega^2)}{\sqrt{m_{e}^{2}-\omega^{2}}},\quad m_{e}^{2}-\omega^{2}>0.
\]
Finally, since the Hamiltonian $H$ is self-adjoint the spectral theorem implies that for every $\psi\in\mathcal{C}_{0}^{\infty}(\Omega)^{4}$
\begin{multline}\label{propagator}
\psi(t,x)=e^{itH}\psi(x)=\int_{\mathbb{R}}d\omega\,e^{i\omega t}\sum_{j\in\mathbb{Z}\backslash\{0\}}\sum_{k\in\mathbb{Z}}\langle\psi^{kj}_{\omega}|\psi_{\omega}\rangle\psi^{kj}_{\omega}(x)+\\
+\sum_{j\in\mathbb{Z}\backslash\{0\}}\sum_{k\in\mathbb{Z}}e^{i\Omega_{k}t}\langle\psi^{kj}_{\Omega_{k}}|\psi_{\Omega_{k}}\rangle\psi^{kj}_{\Omega_k}(x).
\end{multline}
The above expression is the integral representation of the Dirac propagator in an extreme Kerr manifold.

\section*{Acknowledgements}
The research of D.B. was supported by the EU grant HPRN-CT-2002-00277. The authors thank Prof. Gian Michele Graf, Institute for Theoretical Physics, ETH Zurich, Switzerland for his constructive comments during numerous fruitful discussions.

\end{document}